\documentclass[12pt]{article}
\usepackage{amssymb}
\usepackage[pdftex]{graphicx}
\title{ The first  computer program  }
\author{Raul Rojas\\University of Nevada Reno}
\date{March 2023}  
\begin{document}
\maketitle

\begin{abstract}
In 1837, the first computer program  in history was sketched by the renowned mathematician and inventor Charles Babbage. It was a program for  the Analytical Engine. The program consists of a sequence of arithmetical operations and the necessary variable addresses (memory locations)  of the arguments and the result, displayed in tabular fashion, like a program trace. The program computes the  solutions for a system of two linear equations in two unknowns. 
\end{abstract}

\section{Introduction}

This paper describes Charles Babbage’s first computer program, sketched in 1837, that is, almost two hundred years ago. There was no computer  to run the program, only a theoretical concept of what that mechanism would be able to do. Of course, we are talking about the Analytical Engine (AE), the mechanical calculating machine dreamed by the English mathematician and inventor. If it had been built, it would have been the first computer in the world  \cite{bab1864},\cite{bromley}. This was the heyday of the first Industrial Revolution, the age of steam machines and mechanization. Electrical power, light bulbs and telephones were still decades away, but the computer was taking shape in Babbage’s drawing board.

Charles Babbage (1791-1871) developed detailed blueprints for the AE and wrote, between 1836 and 1841,  26 programming examples. The Science Museum in London has digitized the Babbage Archives so that, today, we can inspect the existing diagrams of the Analytical Engine (with first drafts from 1835 on) and the 26 programming examples using a  computer at home. I discuss the architecture of the AE and review some of its programs in \cite{rojas}.

The design of the Analytical Engine consisted of a processor for the four arithmetical operations, called the “mill”, and a separate memory for decimal integers, called the “store” (Babbage once considered building the AE to be able to store hundreds of variables with 40 digits \cite{bab1837}). This separation between processor and memory is typical of computers today. However, the program was not stored in the memory, it was encoded in punched cards, which were read one by one. There was a stream of cards for the processor and a separate stream for the memory (for the addresses of the arguments). In what follows, I sometimes refer to the ``mill" as the processor and to the ``store" as the memory of the machine. When Babbage talks about the contents of memory cells, he 
calls them ``variables". The address of a variable is its subindex. For example, the memory cell with address 3 would be $v_3$.

When the operation punched card for an addition was read, for example,  the mill  transitioned to the “addition” state, while the variable cards would order the store to retrieve and send the contents of the addresses of the two needed arguments to the mill. Reading a variable was a destructive operation, it reset the variable to zero. It was however possible to store the complement of that variable in another memory address, while it was sent to the mill. Rereading that complement and transferring it back to the original memory address (complementing again) restored that address to its original contents.

\section{The first code table}

The program we are discussing  was written by Babbage in 1837. The title of the sketch is``Notations and Calculations" and the first line reads
``No. 1. 4 August 1837." That is, this was the first of the series of programs that Babbage decided to sketch carefully and we even have the date for the program.  The Babbage archive lists this program as "BAB L1". There is a small program for computing a simple formula in the archive, but it is undated and unnumbered (BAB L26). The Babbage archive dates that fragment of code to August 1837, without further information. There is also a sketch about how to assign coefficients of a linear equation to memory addresses, but without code. That fragment's date is given as 1836.

The program in BAB L1  deals with the solution of a system of two linear equations in two variables.  In such case  it is easy to find a closed formula for the result. Babbage considered the two linear equations
$$
ax + by + c = 0
$$
$$
a’x + b’y + c’=0.
$$
We have six parameters, that is, the six coefficients $a,b,c,a’,b’,c’$, present in the two equations. The solution for $x$ is
$$
x=\frac{bc'-b'c}{b'a-ba'}.
$$
Given $x$, the solution for $y$ is
$$
y=(-ax-c)/b
$$
In the first expression, we assume that the denominator $b’a-ba'$ is different from zero, while in the second we assume that $b$ is nonzero, so that the solutions exist. Babbage did not check those two conditions in his program.

First, Babbage  assigns the six coefficients $a,b,c,a’,b’,c’$, to the six variables $v_1$ to $v_6$  in the memory of the AE. He then computes successively the intermediate results $b’a, b'c, ba', bc', bc'-b'c, b’a-ba’ $, and finally the quotient for finding $x$. The complete computation for $x$ requires four multiplications, two subtractions, and a final division. That is grand total of five “big” and two ``small" operations.

Babbage skteched two complete tables for the  computation. The first table is reproduced below.

 \renewcommand{\arraystretch}{2}
 
\hspace*{-2.2cm}{\tiny
\begin{tabular}{ccc|ccccccccccccc}
		 & \textbf{Mill} &   &   &   &   & \textbf{Store} &   &   &   &   &   &   &  &   &  \\
		Numbers of & Nature of & in variables & $+ax$ & $+by$ & $+c$ &  & $+a'x$ & $+b'y$ & $+c'$ &   &   &   &  &   \\  
		 Operations & operations & variables in store & $v_1$ & $v_2$ & $v_3$ &   & $v_4$ & $v_5$ & $v_6$ &   & $v_7$ &   &   &    \\\hline 

		1 & $\times$ & $b'a$ & 0 &   &   &   &   & 0 &   &   & $b'a$ &   & $v_7 =$ & $v_5 \cdot v_1$  \\ 
		2 & $\times$ & $b'c$ & $b'c$ &   & 0 &   &   &   &   &   &   &   & $v_1'=$ & $v_5 \cdot v_3$  \\ 
		3 & $\times$ & $ba'$ &   & 0 & $ba'$ &   & 0 &   &   &   &   &   & $v_3'=$ & $v_2 \cdot v_4$  \\ 
		4 & $\times$ & $bc'$ &   & $bc'$ &   &   &   &   & 0 &   &   &   & $v_2'=$ & $v_2 \cdot v_6$  \\ 
		5 & $-$ & $bc'-b'c$ & 0 & 0 &   &   & $bc'-b'c$ &   &   &   &   &   & $v_4'=$ & $v_2'-v_1'$  \\ 
		6 & $-$ & $b'a-ba'$ & $b'a-ba'$ &   & 0 &   &   &   &   &   & 0 &   & $v_1''=$ & $v_7-v_3'$  \\ 
		7 $*$ & $\div$ & $\frac{bc'-b'c}{b'a-ba'}$ & 0 &   & $=x$ &   & 0 &   &   &   &   & $x=$ & $v_3''=$ & $\frac{v_4'}{v_1''}$  \\ 
		\end{tabular}
		}
		
The table shows the sequence of the seven needed operations. The second column shows the operation and the third column is a comment about the result being computed. The program finishes when the value of  $x$ has been found.

The six memory addresses $v_1,\ldots,v_6$ contain, at the beginning, the coefficients of the six terms $ax,by,c,a'x,b'y,c'$. In the first multiplication the processor uses the variables $v_1$ and $v_5$ for computing $b'a$. The table shows that the two variables have been reduced to zero, after the first operation, and the result $b'a$ is stored in $v_7$. The last column is a comment about the operation which has been performed, i.e., $v_7=v_5\cdot v_1$.

The second multiplication computes $b'c$ and stores it in $v_1$. Symbolically, the computation is $v'_1 = v_5\cdot v_3$. The prime means that the original content of $v_1$ has been overwritten once. However, there is a problem.

Variable $v_5$ was read destructively for the  multiplication in the first row.  The arithmetical operations need two arguments, in this case for the multiplication. 
Babbage designed the AE so that an argument could be reused repetitively. Since the first two computed terms are
$b'a$ and $b'c$, the first argument $b'$ can be retained in the processor. After the multiplication with $a$, we only need to load argument $c$ to the processor. This way of reusing an argument in the processor is not described in BAB L1, but it something that Babbage exploited in other programs.
In BAB L1 Babbage mentions explicitly that $b'$ is reused for a multiplication table with $b'$.

The two columns of comments, displayed side by side, tell the whole story for this computation. In a sense, that is the program that Babbage has in mind, but what the punched cards contain are the specific operations and the addresses needed.

 \renewcommand{\arraystretch}{1.2}
\begin{center}
\begin{tabular}{c|c|ccc}
		 & Computation & Code \\
		\hline 

		1 & $b'a$ &  $v_7 =$ & $v_5 \cdot v_1$ \\ 
		2 &  $b'c$ &  $v_1'=$ & $v_5 \cdot v_3$ \\ 
		3 & $ba'$ &   $v_3'=$ & $v_2 \cdot v_4$  \\ 
		4 & $bc'$ &   $v_2'=$ & $v_2 \cdot v_6$  \\ 
		5 & $bc'-b'c$ &  $v_4'=$ & $v_2'-v_1'$  \\ 
		6 & $b'a-ba'$ &  $v_1''=$ & $v_7-v_3'$  \\ 
		7 &$\frac{bc'-b'c}{b'a-ba'}$ & $v_3''=$ & $\frac{v_4'}{v_1''}$  \\ 
		\end{tabular}
	\end{center}

As can be seen from the code, Babbage reuses memory addresses, and every time a memory address is overwritten he adds a quote to the variable's name. Variables 1 and 3 are reused (overwritten) twice in the program, so that their names become $v''_1$ and $v''_3$.

In other programs written after this first one, Babbage simplified. He not always kept track of variable reuse (with the quotes), since it does not affect the computation. Also, he not always added symbolic comments to the tables, writing only the needed arithmetical operation and the addresses used.

\section{The second code table}

Babbage wrote the second part of the  computation in the same document (L26, kept in the Babbage archive). After having found $x$ with the first seven rows of the program, we can now compute $y$ using the value of $x$. The computation for $x$ is as before,
but $y$ is then computed as $y=( -c -ax)/b$, since the first linear equation is $ax+by+c=0$. The program is reproduced below.

 \renewcommand{\arraystretch}{2}
\hspace*{-4.0cm}{\tiny
		\begin{tabular}{ccc|ccccccccccccccc}
		 	  &\textbf{Mill}&						    &   		& & & \textbf{Store}&   		&   	&   	&   &   	&   	&   	&   &   		& \\
		Numbers of & Nature of & in variables 			& $+ax$ 	& $+by$ & $+c$ 	&	& $+a'x$	& $+b'y$& $+c'$ &   &   	& 		&   	&   &  			&		\\ 
		Operations & operations & variables in store 	& $v_1$ 	& $v_2$ & $v_3$	&	& $v_4$ 	& $v_5$ & $v_6$ &   & $v_7$ & $v_8$ & $v_9$ &   &			&		 \\  \hline
		1 		& $\times$ 	& $b'a$ 					& 0 		&   	&   	&	&   		& 0 	&   	&   & $b'a$	& $Ca$  &		& 	& $v_7 =$ 	& $v_5 \cdot v_1$ 		 \\ 
		2 		& $\times$ 	& $b'c$ 					& $b'c$ 	&   	& 0 	&	&   		& $Cc$  &   	&   &   	&   	&		& 	& $v_1'=$ 	& $v_5 \cdot v_3$ 		 \\ 
		3 		& $\times$ 	& $ba'$ 					&   		& 0 	& $ba'$ &	& 0 		&   	&   	&   &   	&   	& $Cb$	& 	& $v_3'=$ 	& $v_2 \cdot v_4$ 		 \\ 
		4 		& $\times$ 	& $bc'$ 					&   		& $bc'$ &   	&	&   		&   	& 0 	&   &   	&   	&		& 	& $v_2'=$ 	& $v_2 \cdot v_6$ 		 \\ 
		5 		& $-$ 		& $bc'-b'c$ 				& 0 		& 0 	&   	&	& 			&   	&$bc'-b'c$& &   	&   	&		& 	& $v_6'=$ 	& $v_2'-v_1'$ 			 \\ 
		6 		& $-$ 		& $b'a-ba'$ 				& 			&$b'a-ba'$ & 0 	&	&   		&   	&   	&   & 0 	&   	&		& 	& $v_2''=$ 	& $v_7-v_3'$ 			 \\ 
		7 $*$ 	& $\div$ 	& $\frac{bc'-b'c}{b'a-ba'}$ & 	 		& 0  	& 		&	& $=x$ 		&   	& 0  	&   & 	$=x$ 	&		&		&$x=$& $v_4=$ & $\frac{v_6'}{v_2''}$ 	 \\ 
		8		&			&							& $a$		&		&		&	&			&		&		&	&		& 0		&		&	& $v_1''=$	& $v_1 = a$				\\
		9		&			&							& 			& $b$	&		&	&			&		&		&	&		&		& 0		&	& $v_2'''=$	& $v_2 = b$				\\
		10		&			&							&			&		& $c$	&	&			& 0 	&		&	&		&		&		&	& $v_3''=$	& $v_3 = c$				\\
		11		& $\times$	& $ax$						& 0			&		&		&	& 0 		& $ax$	&		&	&		&		&		&	& $v_5''=$	& $v_1'' \cdot v_4'$	\\
		12		& $-$		& $-c-ax$					& $-c-ax$	&		& 0		&	&			& 0 	&		&	&		&		&		&	& $v_1'''=$	& $-v_3''-v_5''$		\\
		13 $*$	& $\div$	& $\frac{-c-ax}{b}$			& 0			& 0		&		&	&			& $=y$	&		&	&		&		&		&$y=$& $v_5'''=$&$\frac{v_1'''}{v_2'''}$\\

		\end{tabular} }

There is something new in the rows 1,2 and 3. Now, Babbage has made explicit that the coefficients $a,b,c$ need to be refreshed, storing their complements $Ca,Cb,Cc$ in auxiliary variables. The complement of $a$ is stored in $v_8$, the complement of $c$ in $v_5$ and the complement of $b$ in $v_9$. We need $a$, $b$ and $c$ for the computation of $y$. 

What Babbage intended to do with the AE, was to store the value of a variable, that is still needed, in another auxiliary variable, when it is sent to the processor. Since the AE used gears (think of a clockface with the digits 0 to 9), storing a number was done rotating the gear anticlockwise, for example, and retrieving the contents was done rotating in the opposite direction until the variable was reduced to zero. Assume, for example, that we have stored the number three, by advancing a gear anticlockwise three positions (out of 10 possible positions, one for each decimal digit). When reading the number, the gear rotates three positions back, in the  clockwise direction. Starting from zero, a receiving auxiliary gear will be rotated  clockwise to the position 7 (the decimal complement of 3). Therefore, the auxiliary variable will not store the original decimal number, but its complement, for every digit of the number. If we had the number 345 in $v_5$, and its complement was saved temporarily in $v_8$, we would have 765 in variable $v_8$. Reading back from $v_8$ to $v_5$, we would complement the number again, digitwise, and $v_5$ would be restored to the original 345.

In the program, the stored complements are transferred back (complementing again) to the variables $v_1$, $v_2$ and $v_3$ in the auxiliary steps 8, 9 and 10. The value of $y$ is computed in steps 11, 12, 13, and the program finally stops.

In other programs, written after this one, Babbage overlapped the storage of the complement of a variable, with its subsequent reading and complementing in a single program step. That is, the store would send a number to the mill, store it temporarily as a complement, and when the result of the computation was delivered by the processor, the stored variable could be refreshed. It is not quite clear if refreshing happened while the processor was busy, or after it had delivered its result to the memory. The notation used by Babbage to indicate that a variable containing the value $a$, for example, kept its value, was $0 /a$, indicating that the variable was reduced to zero and later restored, without necessarily indicating the auxiliary address used.

Symbolically, the complete program written by Babbage would read as follows:

 \renewcommand{\arraystretch}{1.2}
\begin{center}
		\begin{tabular}{c|c|cccccccccccccccc}

		& Computation & Code \\  \hline
		1 		& $b'a$ 					&  $v_7 =$ 	& $v_5 \cdot v_1$ 		 \\ 
		2 		& $b'c$ 					&  $v_1'=$ 	& $v_5 \cdot v_3$ 		 \\ 
		3 		&$ba'$ 					&  $v_3'=$ 	& $v_2 \cdot v_4$ 	 \\ 
		4 		& $bc'$ 					&   $v_2'=$ 	& $v_2 \cdot v_6$ 	 \\ 
		5 		& $bc'-b'c$ 				&  $v_6'=$ & $v_2'-v_1'$ 	 \\ 
		6 		& $b'a-ba'$ 				&  $v_2''=$ & $v_7-v_3'$ 	 \\ 
		7 $*$ 	& $\frac{bc'-b'c}{b'a-ba'}$ & 	 $v_4'=$ & $\frac{v_6'}{v_2''}$ 	 \\ 
		8		&			&			 $v_1''=$	& $v_1 = a$			\\
		9		&			&			 $v_2'''=$	& $v_2 = b$		\\
		10		&			&			 $v_3''=$	& $v_3 = c$		\\
		11		& $ax$				&	$v_5''=$	& $v_1'' \cdot v_4'$	\\
		12		&  $-c-ax$				&	 $v_1'''=$	& $-v_3''-v_5''$		\\
		13 $*$	&  $\frac{-c-ax}{b}$		&	 $v_5'''=$&$\frac{v_1'''}{v_2'''}$& \\

		
		\end{tabular}

\end{center}

In  steps 8, 9 and 10, there is no operation in the processor and only the memory is active, transferring numbers in order to recover the parameters $a,b,c$ from their complements.

\section{Conclusions}

Solving systems of linear equations is very useful in many branches of mathematics and engineering. It is  natural that Babbage
decided to take this as a kind of benchmark problem for the AE. It is known that Chinese mathematicians could solve linear systems of up
to three variables and equations more than two thousand years ago. 

In his code sketches, Babbage did not write high-level code, and then compiled the program. The annotations in his program are more like comments
and the real code would be the strings of punched cards for the processor and for the memory. In later programs, Babbage did not include
a symbolic comment about the computations being performed. Babbage wrote his programs by listing  the needed operations and the needed arguments. Both things translate immediately
 to  the necessary punched cards.  In modern parlance, Babbage wrote his programs in ``assembler". Also, since the operation cards are disconnected from the variable cards, they can synchronize in diverse ways, as explained in \cite{rojas}. Those complications are not present in the program discussed here.

 It is important to point out, although it is obvious, that the first computer program ever written does not belong to  those published in \cite{men1843}.
 That publication appeared six  years after Babbage had already sketched his program ``number one" for solving simultaneous linear equations, and many of the other 25 coding examples.

\end{document}